\documentclass[11pt,reqno]{amsart}
\usepackage{amsfonts}
\usepackage{mathrsfs}
\allowdisplaybreaks[4]
\usepackage[dvips]{graphicx} 
\usepackage{epsfig}
\usepackage{amssymb}
\usepackage{amsmath}
\newcommand{\pa}{\partial}

\begin{document}

\title[Darboux]{ Rogue waves of the Fokas-Lenells equation}
\author{Jingsong He$^{\dag*}$, Shuwei Xu$^{\dag}$, Kuppuswamy Porsezian $^{\ddag}$
 }

\thanks{$^*$Corresonding Author: Email: hejingsong@nbu.edu.cn, Tel: 86-574-87600739, Fax: 86-574-87600744}

 \maketitle
\dedicatory { \dag Department of Mathematics, Ningbo University,
Ningbo, Zhejiang 315211, P.\ R.\ China}\\
\dedicatory { \mbox{\hspace{0.4cm}}\ddag Department of Physics,
Pondicherry University, Puducherry 605014, India }

\begin{abstract}
The Fokas-Lenells (FL) equation arises as a model eqution which
describes for nonlinear pulse propagation in optical fibers by
retaining terms up to the next leading asymptotic order (in the
leading  asymptotic order the  nonlinear Schr\"odinger (NLS)
equation results). Here we present an explicit analytical
representation for the rogue waves of the FL equation. This
representation is constructed by deriving an appropriate Darboux
transformation (DT) and  utilizing a Taylor series expansion of the
associated breather solution. when certain higher-order nonlinear
effects are considered, the propagation of rogue waves in optical
fibers is given.
\end{abstract}

\mbox{\hspace{3cm}} Darboux transformation, breather solution, rogue
wave.

{\bf PACS(2010) numbers}: 02.30.Ik, 42.81.Dp, 52.35.Bj, 52.35.Sb,
94.05.Fg


\section{introduction}

The Fokas-Lenells equation
(FL)\cite{Fokas1,Lenells1,Lenells2,Lenells4,Lenells3}
\begin{equation}\label{FL}
iq_{xt}-iq_{xx}+2q_{x}-q_{x}qq^{\ast}+iq=0,
\end{equation}
is one of the important models from both mathematical and  physical
considerations. In eq.(\ref{FL}), $q$ represents complex field
envelope and asterisk denotes complex conjugation, and subscript $x$
(or $t$) denotes partial derivative with respect to $x$ (or $t$).
The FL equation \cite{Fokas1,Lenells2} is related to the nonlinear
Schr\"odinger (NLS) equation in the same way that the Camassa-Holm
equation is associated with the KdV equation.  In optics,
considering suitable higher order linear and nonlinear optical
effects,  the FL equation has been derived as a model to describe
femtosecond pulse propagation through single mode optical silica
fiber and several interesting solutions have been constrcuted
\cite{Lenells3}.

Here, considering the physical significance of the FL equation and
inspired by the importance of the recent interesting developments in
the analysis of rogue waves of the NLS and the DNLS equations, we
shall construct rogue wave solutions of the FL equation, by using
the Darboux transformation
(DT)\cite{GN,matveev,he,kenji1,steduel,xuhe}. Our construction
reveals that there exists a difference between the FL system and
other integrable models, like the Ablowitz-Kaup-Newell-Segur (AKNS)
\cite{AKNS,Ablowitz} system and the Kaup-Newell (KN)
\cite{Ablowitz,KN} system.

Before discussing the DT of the FL system, let us briefly discuss
the importance of rogue waves in mathematics and physics. Rogue
waves have recently been the subject of intensive investigations in
oceanography\cite{Kharif1,Kharif2,Akhmediev1,Ankiewicz,Akhmediev2,
Didenkulova}, where they occur due to modulation instability
\cite{Peregrine,Dysthe1,Zakharov1,Zakharov2,Akhmediev3,MaB,Crespo,yan,guo}
, either through random initial condition \cite{Didenkulova,Ying} or
via other processes \cite{Akhmediev1,Ankiewicz,Akhmediev2,
Peregrine,Dysthe1,Zakharov1,
Zakharov2,MaB,Akhmediev3,Crespo,yan,guo,jianke,lai}. The first order
rogue wave usually  takes the form of a single peak hump with two
caves in a plane with a nonzero boundary. One of the possible
generating mechanisms for  rogue waves is through the creation of
breathers\cite{Akhmediev1,Ankiewicz,Akhmediev2,Peregrine,
Dysthe1,Zakharov1,Akhmediev3,MaB,Crespo,guo,jianke} which can be
formed due to modulation instability.  Then, larger rogue waves can
emerge when two or more breathers collide with each other
\cite{Akhmediev4}. Rogue waves have  also been observed in space
plasmas\cite{xuhe,ruderman,derman,Moslem}, as well as in optics when
propagating high power optical radiation through photonic crystal
fibers \cite{Optical1,Optical2,Optical3}. Considering higher order
effects in the propagation of femtosecond pulses, rogue waves have
been reported in the Hirota equation and the NLS-MB
system\cite{Akhmediev5,taohe1,hexuk}.

This paper is organized as follows. Section 2, presents a simple
approach to DT for the FL system, the determinant representation of
the 1-fold DT,  and  formulae of $q^{[1]}$ and $r^{[1]}$ expressed
in terms of eigenfunctions of the associated spectral problem. The
reduction of DT to the FL equation is also discussed by choosing
appropriate pairs of eigenvalues and eigenfunctions. Section 3
presents the construction of rogue wave solutions by using a Taylor
series expansion about the breather solution generated by DT from a
periodic seed solution with constant amplitude. Finally, in section
4 we conclude our results.

\section{Darboux transformation}
Let us start from the non-trivial flow of the FL (Fokas-Lenells)
system\cite{Lenells2} in the following modified form,
\begin{equation}\label{sy1}
iq_{xt}-iq_{xx}+2q_{x}-q_{x}qr+iq=0,
\end{equation}
\begin{equation}\label{sy2}
ir_{xt}-ir_{xx}-2r_{x}+r_{x}rq+ir=0,
\end{equation}
which are  exactly reduced to the FL eq.(\ref{FL}) for $r=q^\ast$
while the choice $r =-q^\ast$ would lead to eq.(\ref{FL}) with
change in sign of the nonlinear term. The Lax pairs corresponding to
coupled FL eq.(\ref{sy1}) and (\ref{sy2}) can be given by the FL
spectral problem of the form \cite{Lenells2}
\begin{equation}\label{sys11}
 \pa_{x}\psi=(J\lambda^2+Q\lambda)\psi=U\psi,
\end{equation}
\begin{equation}\label{sys22}
\pa_{t}\psi=(J\lambda^2+Q\lambda+V_{0}+V_{-1}\lambda^{-1}+\dfrac{1}{4}J\lambda^{-2})\psi=V\psi,
\end{equation}
with
\begin{equation}\label{fj1}
    \psi=\left( \begin{array}{c}
      \phi \\
      \varphi\\
     \end{array} \right),\nonumber\\
   \quad J= \left( \begin{array}{cc}
      -i &0 \\
      0 &i\\
   \end{array} \right),\nonumber\\
  \quad Q=\left( \begin{array}{cc}
     0 &q_{x} \\
     r_{x} &0\\
  \end{array} \right),\nonumber\\
\end{equation}
\begin{equation}\label{fj2}
V_{0}=\left( \begin{array}{cc}
i-\frac{1}{2}iqr &0 \\
 0&-i+\frac{1}{2}iqr\\
\end{array} \right),\quad V_{-1}=\left( \begin{array}{cc}
0 &\frac{1}{2}iq \\
- \frac{1}{2}ir&0\\
\end{array} \right).\nonumber\\
\end{equation}
Here $\lambda$, an arbitrary complex number, is  called the
eigenvalue(or isospectral parameter), and $\psi$ is called the
eigenfunction associated with  $\lambda$ of the FL system. Equations
(\ref{sy1}) and (\ref{sy2}) are equivalent to the compatibility
condition  $U_{t}-V_{x}+[U,V]=0$ of (\ref{sys11}) and (\ref{sys22}).

Now, let us consider a matrix $T $ of gauge transformation for the
spectral problem (\ref{sys11}) and (\ref{sys22}) with the following
form
\begin{equation}
\label{bh3} \psi^{[1]}=T~\psi,
\end{equation}
and then
  \begin{equation}\label{bh1}
{\psi^{[1]}}_{x}=U^{[1]}~\psi^{[1]},\ \ U^{[1]}=(T_{x}+T~U)T^{-1}.
\end{equation}
\begin{equation}\label{bh2}
{\psi^{[1]}}_{t}=V^{[1]}~\psi^{[1]}, \ \ V^{[1]}=(T_{t}+T~V)T^{-1}.
\end{equation}\\
By cross differentiating (\ref{bh1}) and (\ref{bh2}), we obtain
\begin{equation}\label{bh4}
{U^{[1]}}_{t}-{V^{[1]}}_{x}+[{U^{[1]}},{V^{[1]}}]=T(U_{t}-V_{x}+[U,V])T^{-1}.
\end{equation}
This implies that, in order to prove eqs.(\ref{sy1}) and (\ref{sy2})
are invariant under the transformation (\ref{bh3}), it is crucial to
construct a matrix $T$ so that $U^{[1]}$ and $V^{[1]}$  have the
same forms as that of  $U$ and $V$. At the same time the old
potentials (or seed solutions)($q$, $r$) in spectral matrixes $U$
and $V$ are mapped into new potentials (or new solutions)($q^{[1]}$,
$r^{[1]}$) in terms of  transformed spectral matrixes $U^{[1]}$ and
$V^{[1]}$. This newly obtained matrix $T$ is a Darboux
transformation of the FL system of eqs. (\ref{sy1}) and (\ref{sy2}).

Based on the DT for the NLS\cite{GN,matveev,he} and the
DNLS\cite{kenji1,steduel,xuhe}, we suppose that a trial Darboux
matrix $T$ in eq.(\ref{bh3}) is assumed to be  in the form
\begin{equation}\label{TT}
 T_{1}=T_{1}(\lambda;\lambda_1;\lambda_2)=\left( \begin{array}{cc}
a_{1}&0\\
0 &d_{1}\\
\end{array} \right)\lambda+\left( \begin{array}{cc}
0&b_{0}\\
c_{0} &0\\
\end{array} \right)+\left( \begin{array}{cc}
a_{-1}&0\\
0&d_{-1}\\
\end{array} \right)\lambda^{-1}.
\end{equation}
Here $a_{1}, d_{1},b_{0},c_{0}$ are undetermined coefficients and
they are function of ($x$, $t$), which will be expressed in terms of
the eigenfunctions associated with
 $\lambda_1$ and $\lambda_2$ in the FL spectral problem
 and $a_{-1}$ and
 $d_{-1}$ are constants. For our further analysis, setting two eigenfunctions $\psi_j$ as
\begin{eqnarray}
&&\psi_{j}=\left(
\begin{array}{c}\label{jie2}
 \phi_{j}   \\
 \varphi_{j}  \\
\end{array} \right),\ \ j=1,2,\phi_{j}=\phi_{j}(x,t,\lambda_{j}), \
\varphi_{j}=\varphi_{j}(x,t,\lambda_{j}). \label{jie1}
\end{eqnarray}
After tedious but straightforward calculations, we get the one-fold
Darboux transformation of the FL system as
 \begin{eqnarray}\label{onefoldDT}
T_1(\lambda;\lambda_1;\lambda_2)=\left(
\begin{array}{cc}
a_{1}\lambda +a_{-1}\lambda^{-1}& b_{0}\\
c_{0} & d_{1}\lambda +d_{-1}\lambda^{-1}
\end{array}  \right),  \label{DT1matrix}
\end{eqnarray}
and then the corresponding new solutions $q^{[1]}$ and $r^{[1]}$ are
given by
\begin{eqnarray}\label{sTT}
q^{[1]}=q \dfrac{a_{-1}}{d_{-1}}+\dfrac{b_{0}}{d_{-1}}, r^{[1]}=r
\dfrac{d_{-1}}{a_{-1}}+\dfrac{c_{0}}{a_{-1}}.
\end{eqnarray}
With the condition  $a_{-1}=d_{-1}=1$ and
\begin{eqnarray}
a_{1}= \dfrac{\begin{vmatrix}
-{\lambda_{1}}^{-1}\phi_{1}&\varphi_{1}\\
-{\lambda_{2}}^{-1}\phi_{2}&\varphi_{2}\nonumber\\
\end{vmatrix}}{\begin{vmatrix}
\lambda_{1}\phi_{1}&\varphi_{1}\\
\lambda_{2}\phi_{2}&\varphi_{2}\nonumber\\
\end{vmatrix}},\quad
d_{1}= \dfrac{\begin{vmatrix}
-{\lambda_{1}}^{-1}\varphi_{1}&\phi_{1}\\
-{\lambda_{2}}^{-1}\varphi_{2}&\phi_{2}\nonumber\\
\end{vmatrix}}{\begin{vmatrix}
\lambda_{1}\varphi_{1}&\phi_{1}\\
\lambda_{2}\varphi_{2}&\phi_{2}\nonumber\\
\end{vmatrix}},\quad
b_{0}= \dfrac{\begin{vmatrix}
\lambda_{1}\phi_{1}&-{\lambda_{1}}^{-1}\phi_{1}\\
\lambda_{2}\phi_{2}&-{\lambda_{2}}^{-1}\phi_{2}\nonumber\\
\end{vmatrix}}{\begin{vmatrix}
\lambda_{1}\phi_{1}&\varphi_{1}\\
\lambda_{2}\phi_{2}&\varphi_{2}\nonumber\\
\end{vmatrix}},\quad
c_{0}= \dfrac{\begin{vmatrix}
\lambda_{1}\varphi_{1}&-{\lambda_{1}}^{-1}\varphi_{1}\\
\lambda_{2}\varphi_{2}&-{\lambda_{2}}^{-1}\varphi_{2}\nonumber\\
\end{vmatrix}}{\begin{vmatrix}
\lambda_{1}\varphi_{1}&\phi_{1}\\
\lambda_{2}\varphi_{2}&\phi_{2}\nonumber
\end{vmatrix}}
\end{eqnarray}

 Having obtained the values of all the unknown coefficients, we are now in a position to consider the reduction of the DT so that $q^{[1]}=(r^{[1]})^*$, then the DT of the FL equation is given. Under the reduction condition $q=r^*$ with
 $\psi_k=\left( \begin{array}{c}
\phi_k\\
\varphi_k
\end{array} \right)$
is an eigenfunction associated with the eigenvalue $\lambda_k$, then
${\lambda_{l}}=\lambda_{k}^*(k\neq l)$ and
$\phi_{l}={\varphi_{k}}^{\ast},\varphi_{l}={\phi_{k}}^{\ast} $. It
is easy to check $q^{[1]}=(r^{[1]})^*$  with the help of following
choice of eigenvalues and eigenfunctions:
\begin{equation}\label{2nfoldredu}
 \lambda_1 \leftrightarrow \psi_1=\left( \begin{array}{c}
\phi_1\\
\varphi_1\end{array} \right) , \text{and} \lambda_{2}=
\lambda_{1}^*,\leftrightarrow \psi_{2}=\left( \begin{array}{c}
\varphi_{1}^*\\
\phi_{1}^*
\end{array} \right).
\end{equation}
Thus, under the above choice of eigenfunctions and eigenvalues, the
resulting $T_1$ in eq.(\ref{onefoldDT}) can be called as one-fold
Darboux transformation of the FL equation. Further, taking one Pair
of eigenvalues $\lambda_{1}=\alpha_{1}+i\beta_{1}$ and
$\lambda_{2}=\alpha_{1}-i\beta_{1}$ and their corresponding
eigenfunctions, the final form of one-fold DT eq.(\ref{onefoldDT})
reduces to a  new solution of the form
\begin{equation}\label{q2j2}
 q^{[1]}=q+\dfrac{({\lambda_{2}}^{2}-{\lambda_{1}}^{2})
 \phi_{1}{\varphi_{1}}^{\ast}}{\lambda_2\lambda_1(-\lambda_{2}\varphi_{1}
 {\varphi_{1}}^{\ast}+\lambda_{1}\phi_{1}{\phi_{1}}^{\ast})},
\end{equation}
Notice that the denominator of $q^{[1]}$  has a nonzero imaginary
part if $\beta_1$ is not zero, which clearly indicates that the new
solution $q^{[1]}$ is a non-singular solution.

\section{rogue waves}
Using the results of DT constructed above, breather solutions of FL
equation can be generated by assuming a periodic seed solution of
eq.(\ref{q2j2}), then we can construct the  rogue waves of the FL
equation from a Taylor series expansion of  the breather solutions.

For this purpose, assuming $a$ and  $c$ as two real constants, then
$q=ce^{(i(a x+(\dfrac{(a+1)^2}{a}-c^2)t))}$ is a periodic solution
of the FL equation, which will be used as a seed solution of the DT.
Substituting the above form of $q$ into the spectral problem
eq.(\ref{sys11})  and eq.(\ref{sys22}), and using the method of
separation of variables and the superposition principle, the
eigenfunction $(\phi_1,\varphi_1)$ associated with $\lambda_1$ are
given explicitly. For simplicity, setting
$c=-\dfrac{\sqrt{a+2{\alpha_1}^{2}-2{\beta_1}^{2}}}{a}$, taking
$(\phi_{1},\varphi_1)$ back into eq.(\ref{q2j2}), then the final
form of the breather solution is obtained in the form
\begin{eqnarray}\label{q2j3}
&&q^{[1]}=\lefteqn{(\dfrac{-2\alpha_{1}\beta_{1}(w_1\cosh(H_1)+w_2sinh(H_1)+w_3\cos(H_2)+w_4sin(H_2))}{w_5\cosh(H_1)+w_6sinh(H_1)+w_7\cos(H_2)+w_8sin(H_2)}{}}\nonumber\\
&&{}\mbox{\hspace{0.5cm}}-\dfrac{\sqrt{a+2{\alpha_1}^{2}-2{\beta_1}^{2}}}{a})\exp(i(ax+(a+2-\dfrac{2({\beta_{1}}^{2}-{\alpha_{1}}^{2})t}{a^2}))),
\end{eqnarray}
where
\begin{eqnarray*}
&&w_1=h_1(-h_3+2\sqrt{h_3}\alpha_{1}),w_2=i\sqrt{-h_1h_2}(-h_3+2\sqrt{h_3}\alpha_{1}),\\
&&w_3=2\alpha_{1}h_1(-2\alpha_{1}+\sqrt{h_3}),w_4=-2i\sqrt{-h_1h_2}\beta_{1}(-2\alpha_{1}+\sqrt{h_3}),\\
&&w_5=2\alpha_{1}\beta_{1}({\alpha_{1}}^{2}+{\beta_{1}}^{2})h_1(-2\alpha_{1}+\sqrt{h_3}),w_6=-2i\alpha_{1}\beta_{1}({\alpha_{1}}^{2}+{\beta_{1}}^{2})\sqrt{-h_1h_2}(-2\alpha_{1}+\sqrt{h_3}),\\
&&w_7=\beta_{1}({\alpha_{1}}^{2}+{\beta_{1}}^{2})h_1(-h_3+2\sqrt{h_3}\alpha_{1}),w_8=-i\alpha_{1}({\alpha_{1}}^{2}+{\beta_{1}}^{2})\sqrt{-h_1h_2}(-h_3+2\sqrt{h_3}\alpha_{1}),\\
&&H_1=\dfrac{\sqrt{-h_1h_2}}{2a({\alpha_{1}}^{2}+{\beta_{1}}^{2})^{2}}((2a({\alpha_{1}}^{2}+{\beta_{1}}^{2})^{2}+({\alpha_{1}}^{2}-{\beta_{1}}^{2}))t+2a({\alpha_{1}}^{2}+{\beta_{1}}^{2})^{2}x),\\
&&H_2=\dfrac{\sqrt{-h_1h_2}}{a({\alpha_{1}}^{2}+{\beta_{1}}^{2})^{2}}\alpha_{1}\beta_{1}t, h_1=a+2({\alpha_{1}}^{2}+{\beta_{1}}^{2}),\\
&&h_2=a-2({\alpha_{1}}^{2}+{\beta_{1}}^{2}),h_3=a+2({\alpha_{1}}^{2}-{\beta_{1}}^{2}).
\end{eqnarray*}
From eq.(\ref{q2j3}), we can easily get the Ma
breathers\cite{MaB}(time periodic breather solution) and the
Akhmediev breathers \cite{Ankiewicz,Akhmediev2,Akhmediev3} (space
periodic breather solution) solution. In general, the solution in
eq.(\ref{q2j3}) evolves periodically along the  straight line with a
certain angle between $x$ and $t$ axis. The dynamical evolution of
$|q^{[1]}|^2$ in eq.(\ref{q2j3}) is plotted in Figure 1.

\begin{figure}[ht]
\setlength{\unitlength}{0.1cm}
\epsfig{file=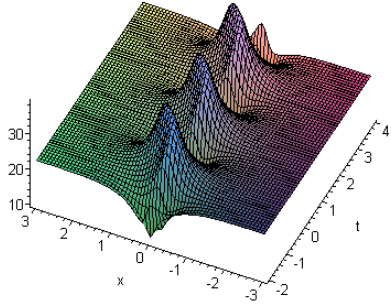,width=10cm}\vspace{-1cm}\caption{{Dynamical
evolution of $|q^{[1]}|^2$({\bf time periodic} breather) with
specific parameters
$\alpha_{1}=1,\beta_{1}=\dfrac{1}{2},a=-\dfrac{6}{25}$. The
trajectory is a line at $x=0$.}}
\end{figure}

 Finally, a Taylor series expansion about the breather solution of the FL equation is used to construct the rogue wave of the FL equation in the following way. By letting $ a \rightarrow 2({\alpha_{1}}^{2}+{\beta_{1}}^{2})$ in
 $(\ref{q2j3})$, we obtain the  rogue wave solution in the form
\begin{equation}\label{focas}
{q}^{[1]}_{rw}=\alpha_{1} \dfrac{s_2}{s_3}\exp{s_1}
\end{equation}
with
\begin{eqnarray*}
&&s_1=i(2 x s+\dfrac{1}{2s}t(4 s^2-2\frac{-{\beta_{1}}^2+s}{s}+4 s+1)),\\
&&s_2=\lefteqn{(t^2(-{\beta_{1}}^2+s) k_1+16 x^2 (-{\beta_{1}}^2+s)
s^4{}}\nonumber\\&&{} \mbox{\hspace{0.8cm}}+8 t x (-{\beta_{1}}^2+s)
k_2 s+2 i t(-{\beta_{1}}^2
+s) (12 s^2+1) s+24 i x (-{\beta_{1}}^2+s) s^3-3 s^3),\\
&&s_3=\lefteqn{(t^2(-{\beta_{1}}^2+s) k_1+16
x^2(-{\beta_{1}}^2+s)s^4{}}\nonumber\\&&{} \mbox{\hspace{0.8cm}}+8 t
x (-{\beta_{1}}^2+s) k_2 s-2it(-{\beta_{1}}^2+s)(4 s^2-1
)s-8ix(-{\beta_{1}}^2+s)s^3+s^3)s,\\
&&k_1=\lefteqn{64(-{\beta_{1}}^4+s{\beta_{1}}^2)(2{\beta_{1}}^{4}-2s{\beta_{1}}^{2}+s^2)+128{\beta_{1}}^{8}-256s{\beta_{1}}^
{6}{}}\nonumber\\
&&{}\mbox{\hspace{0.8cm}}+192s^2{\beta_{1}}^{4}+(-64s^3-16s){\beta_{1}}^{2}+8s^2+16s^4+1,\\
&&k_2=12(-{\beta_{1}}^{4}+s{\beta_{1}}^{2})s+12s{\beta_{1}}^{4}+(-12s^2-2){\beta_{1}}^{2}+s+4s^3,\\
&&s={\alpha_{1}}^2+{\beta_{1}}^{2}.\\
\end{eqnarray*}

By letting $ x \rightarrow {\infty}, \ t \rightarrow {\infty}$,
 so $|{q}^{[1]}_{rw}|^{2}\rightarrow $$\dfrac{{\alpha_{1}}^2}{({\alpha_{1}}^2+{\beta_{1}}^{2})^{2}}$.
The maximum amplitude of $|{q}^{[1]}_{rw}|^{2}$ occurs at $t = 0$
and $x=0$ and is equal to
$\dfrac{9{\alpha_{1}}^2}{({\alpha_{1}}^2+{\beta_{1}}^{2})^{2}}$,and
the minimum amplitude of $|{q}^{[1]}_{rw}|^{2}$ occurs at $t =
\pm\dfrac{9({\alpha_{1}}^{2}+{\beta_{1}}^{2})^2}{2\sqrt{3({\alpha_{1}}^{2}+4{\beta_{1}}^{2})}{\alpha_{1}}}$
and
$x=\mp\dfrac{3(12({\alpha_{1}}^{2}+{\beta_{1}}^{2})^2+1)}{8\sqrt{3({\alpha_{1}}^{2}+4{\beta_{1}}^{2})}{\alpha_{1}}}$
and is equal to $0$. Through Figure 2 and Figure 3 of
$|{q}^{[1]}_{rw}|^2$,  the main features (such as large amplitude
and local property on (x-t) plane) of the rogue wave are clearly
shown. \vspace{1cm}
\begin{figure}[ht]
\setlength{\unitlength}{0.1cm}
\epsfig{file=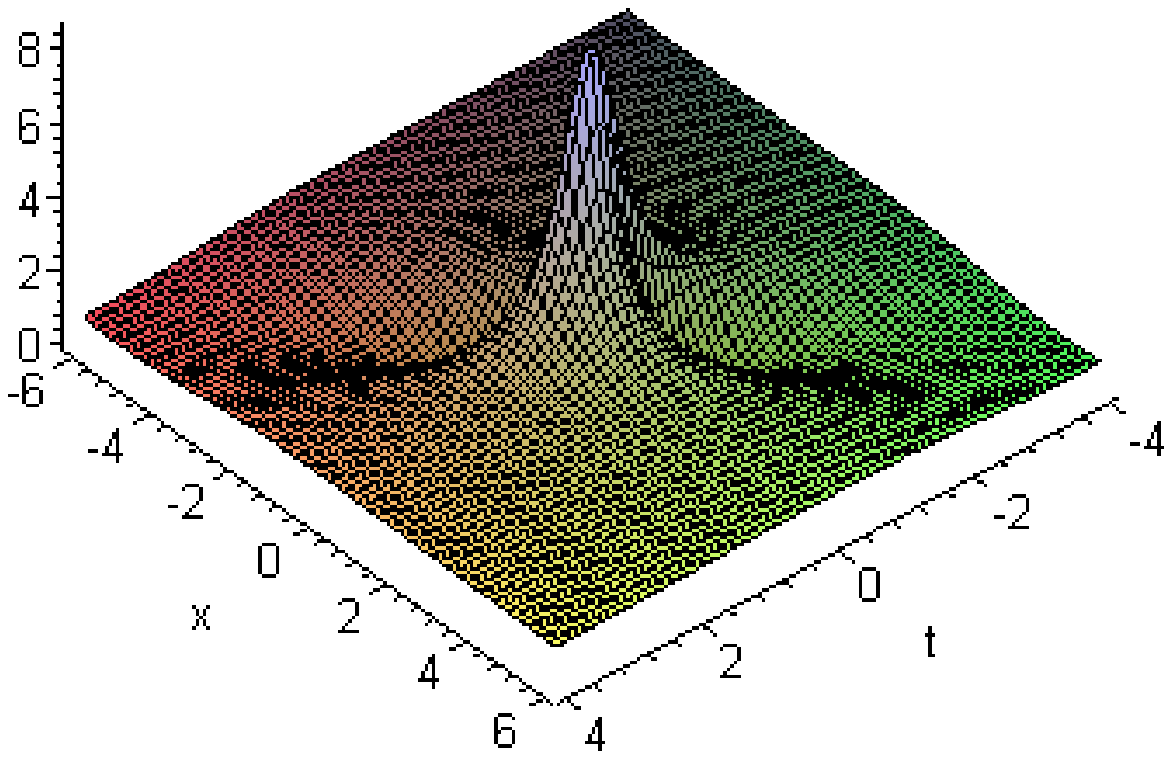,width=12cm}\vspace{-3cm}
\caption{{Dynamical evolution of $|{q}^{[1]}_{rw}|^{2}$ with
specific parameters
$\alpha_{1}=\dfrac{1}{2},\beta_{1}=\dfrac{1}{2}$, in the limit$ x
\rightarrow {\infty}, \ t \rightarrow {\infty}$, so
$|{q}^{[1]}_{rw}|^{2}\rightarrow 1$. The maximum amplitude of
$|q^{[1]}_{rw}|^{2}$ occurs at $t = 0$ and $x=0$ and is equal to
9,and the minimum amplitude of $|q^{[1]}_{rw}|^{2}$ occurs at $t =
\pm\dfrac{3\sqrt{15}}{10}$ and $x=\mp\dfrac{2\sqrt{15}}{5}$ and is
equal to $0$. }}
\epsfig{file=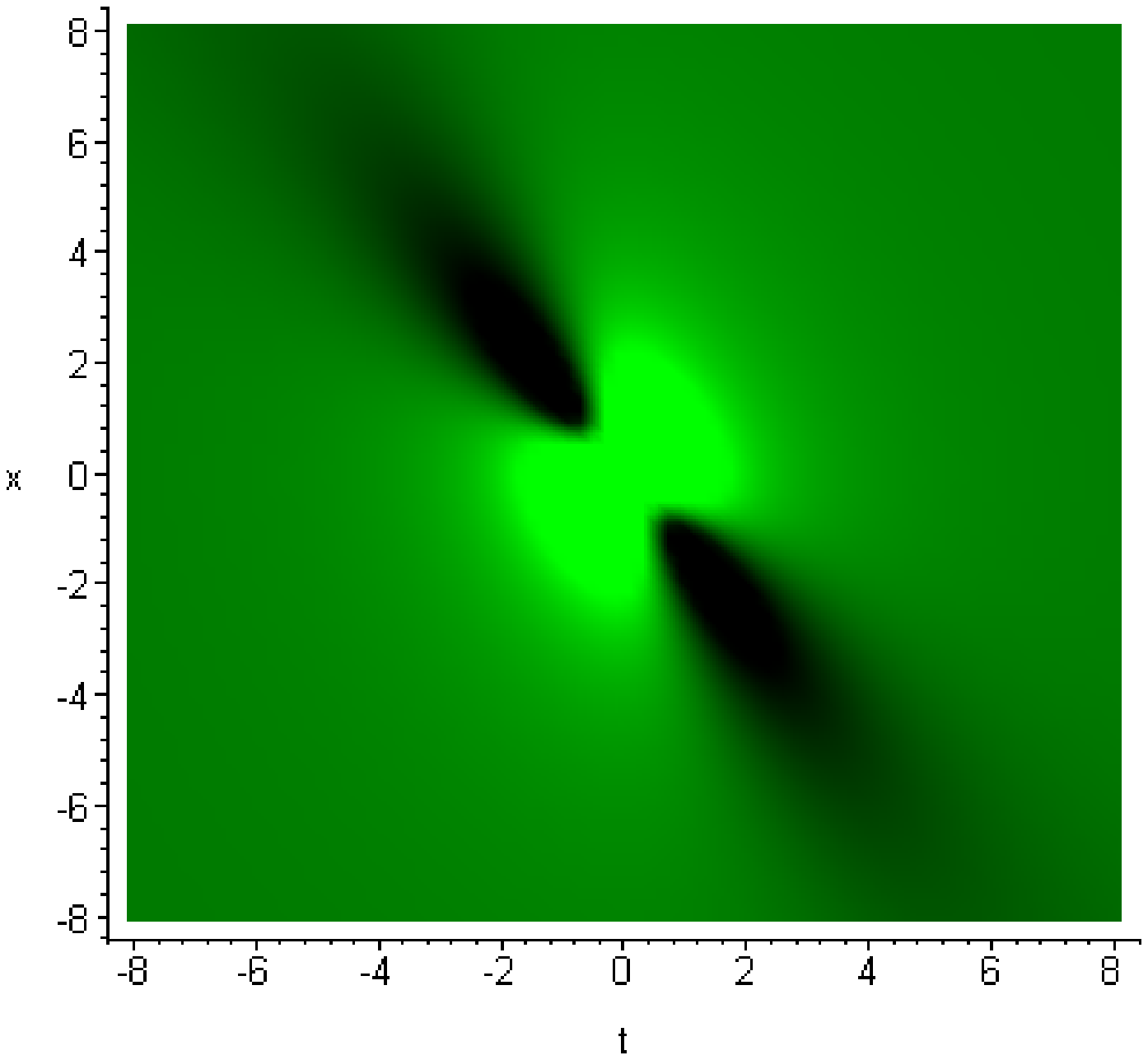,width=8cm} \vspace{-1cm} \caption{{Contour
plot of the wave amplitudes of $|{q}^{[1]}_{rw}|^{2}$ with
$\alpha_{1}=\dfrac{1}{2},\beta_{1}=\dfrac{1}{2}$.}}
\end{figure}

\section{Conclusions}

Thus, in this paper, considering FL system of equation which
describes nonlinear pulse propagation through single mode optical
fiber, the determinant representation of the one-fold DT for the FL
system is given eqs.(\ref{onefoldDT}) and (\ref{sTT}). By choosing
pair of eigenvalues in the form $\lambda_2=\lambda_1^*$ and assuming
suitable seed solutions, the breather solution  and rogue waves of
the FL equation are derived in eq. (\ref{q2j3}) and eq.
(\ref{focas}).  Our results provide an alternative possibility to
observe rogue waves in optical system. From the one-fold DT, it is
interesting to observe that the DT of the FL system exhibits the
following novelty in comparison with other integrable models like
the AKNS and the KN systems: the DT matrix of the FL system has
three different terms depending on $\lambda$.

Thus, the DT as well as the rogue wave of the FL system present
novel features, in comparison with the DT and rogue wave solutions
of the standard integrable systems like the AKNS and the KN systems.
The construction of higher order rogue wave of the FL equation by
using the determinant representation of the DT will be published
elsewhere.

{{\bf Acknowledgments} {\noindent \small This work is supported by
the NSF of China under Grant No.10971109 and K.C.Wong Magna Fund in
Ningbo University. Jingsong He is also supported by Program for NCET
under Grant No.NCET-08-0515 and Natural Science Foundation of Ningbo
under Grant No.2011A610179. We thank Prof. Yishen Li(USTC,Hefei,
China) for his useful suggestions on the rogue wave. J. He thank
Prof. A.S.Fokas and Dr. Dionyssis Mantzavino(Cambridge University)
for many helps on this paper. KP wishes to thank the DST, DAE-BRNS,
UGC, CSIR, Government of India, for the financial support through
major projects.}}

\end{document}